\documentclass{aa}  
 \usepackage[varg]{txfonts}

\usepackage{orcidlink}

\usepackage{natbib}
\bibpunct{(}{)}{;}{a}{}{,}
\usepackage{graphicx}
\usepackage{txfonts}
\usepackage{url}
\usepackage{hyperref}
\hypersetup{
     colorlinks=true, 
     linkcolor=red,  
     filecolor=magenta,   
     citecolor=blue,   
     urlcolor=blue
     }
\begin{document}

   \title{Extended Main-Sequence Turnoff and Red Clump in intermediate-age star clusters: A study of NGC\,419}


   \author{F. Dresbach\inst{1}\fnmsep\inst{2}\orcidlink{0000-0003-0808-8038}
          \and
          D. Massari\inst{3}\orcidlink{0000-0001-8892-4301}
          \and
          B. Lanzoni\inst{1}\fnmsep\inst{3}\orcidlink{0000-0001-5613-4938}
          \and
          F. R. Ferraro\inst{1}\fnmsep\inst{3}\orcidlink{0000-0002-2165-8528}
          \and
          E. Dalessandro\inst{3}\orcidlink{0000-0003-4237-4601}
          \and
          M. Libralato\inst{4}\orcidlink{ 0000-0001-9673-7397}
          \and
          S. Raso\orcidlink{0000-0003-4592-1236}
          }

   \institute{Dept. of Physics and Astronomy, University of Bologna, Via Gobetti 93/2, Bologna, Italy\
         \and Lennard-Jones Laboratories, School of Chemical and Physical Sciences, Keele University, ST5 5BG, UK \\
         \email{f.dresbach@keele.ac.uk}\
        \and INAF - Osservatorio di Astrofisica e Scienza dello Spazio di Bologna, Via Gobetti 93/3, I-40129 Bologna, Italy\
        \and AURA for the European Space Agency (ESA), Space Telescope Science Institute, 3700 San Martin Drive, Baltimore, MD 21218, USA
    }

   \date{}

 
  \abstract{With the goal of untangling the origin of extended main-sequence turnoffs (eMSTOs) and extended red clumps (eRCs) in star clusters, in this work we present the study of the intermediate-age cluster NGC~419, situated along the Bridge of the Small Magellanic Cloud. To this aim, we analyzed multi-epoch, high angular resolution observations acquired with the Hubble Space Telescope for this dynamically young cluster, which enabled the determination of precise proper motions and therefore the assessment of the cluster membership for each individual star in the field of view. With this unprecedented information at hand, we first studied the radial distribution of kinematically selected member stars in different eMSTO subregions. The absence of segregation supports the rotation scenario as the cause for the turnoff color extension and disfavors the presence of a prolonged period of star formation in the cluster. A similar analysis on the eRC of NGC~419 confirms the absence of segregation, providing further evidence against an age spread, which is at odds with previous investigations. Even so, the currently available evolutionary models including stellar rotation fail at reproducing the two photometric features simultaneously. We argue that either shortcomings in these models or a different origin for the red clump feature, such as a nonstandard differential mass loss along the red giant branch phase, are the only way to reconcile our observational findings with theoretical expectations.}


   \maketitle
%

\section{Introduction}\label{sec:intro}

    Globular clusters (GCs) have long been considered the perfect example of simple stellar populations, which are systems composed of stars of uniform ages and chemical compositions. However, in the last decades, spectroscopic and photometric studies have revealed a much more complex situation. On the one hand, a few GC-like stellar systems have been found to harbor multi-iron and multiage subpopulations (see the cases of Omega Centauri in the Galactic halo, and Terzan 5 and Liller 1 in the bulge; e.g., \citealt{2010johnson}, \citealt{2000pancino}, \citealt{Bellini2017a}, \citealt{2009ferraroTerzan}, \citeyear{2016Ferraro}, \citeyear{2021ferraro}, \citealt{2011origlia}, \citeyear{origlia2013}, \citealt{2014massari}, \citealt{2022dalessandro}, \citealt{2023crociati}). On the other hand, light-element abundance variations and splitting of sequences in the color-magnitude diagrams (CMDs) have also been observed in mono-metallic GCs, thus pointing toward much more complex formation scenarios (see \citealt{2019A&ARvGratton} for a review). Furthermore, it has been discovered that many extragalactic clusters of a young and intermediate age also present unusual structures in their observed CMDs, such as a split main sequence (MS) and extended MS turnoff (eMSTO; \citealt{2008Glatt}; \citealt{2009Milone}, \citeyear{2016Milone}, \citeyear{2023milone} and references therein). These features are not predicted by the standard models of stellar evolution, so their discovery increasingly challenged our understanding of how star clusters formed and evolved. In particular, studies of clusters in the Magellanic Clouds (MCs) showed that the eMSTO appears to be a common feature for systems younger than 2 Gyr and a few different scenarios have been proposed to explain this extension, but its origin is still debated (\citealt{10.1111/j.1365-2966.2007.11915.x}). One scenario suggests that the spread in the color of turnoff (TO) stars could be caused by different stellar rotation velocities of massive stars ($1.2-1.7 \ \rm{M_{\odot}}$), which can shape the CMD of a cluster \citep{10.1111/j.1745-3933.2009.00696.x}. The authors indeed pointed out how stars with rapid rotation would have a reduced surface gravity, resulting in lower luminosities and an effective temperature, hence they could be responsible for the fainter and redder extension of the TO. Then, as these massive stars evolve off the MS, they expand and slow their rotation rate down due to the conservation of angular momentum. High rotation is also not expected for lighter stars (below $1.2 \ \rm{M_{\odot}}$), where the presence of a convective envelope may generate a magnetic field, whose braking torque can efficiently slow the rotation rate down (\citealt{1962Schatzman}; \citealt{1987Mesten}). This phenomenon would explain why eMSTOs are not found in old clusters, where all massive stars have long left the MS. Support to this scenario comes from recent photometric and spectorscopic studies of young MC clusters (t$\sim$ 80-400 Myr) that highlighted the presence of eMSTO and split MS ascribed to different rotational velocities (e.g., \citealt{2016Milone}, \citealt{2017Dupreee}, \citealt{2018Marino}). \\
    However, alternative scenarios have been proposed, arguing that other consequences of rotation, such as internal mixing and the magnitude dependence on the star orientation, could have opposite effects in shaping the CMDs (\citealt{2011GirardiEggenbergerMiglio}). Many authors indeed ascribe the extension of the TO to a spread in age among its stars, due to a prolonged period of star formation (\citealt{2008MackeyBroNie}; \citealt{2009Milone}; \citealt{Rubele2010}; \citealt{girardi2013}), so the brighter and bluer region of the TO should be populated by younger and more massive stars. With this interpretation, clusters with an eMSTO could represent the younger counterparts of the old GCs with multiple stellar populations (e.g., \citealt{2011Conroy}). However, both of these scenarios pose some problems and are not able to fully explain the observed morphology of the cluster CMDs (\citealt{2016Milone}; \citealt{2017Dantona}). For the rotation scenario, it still remains unclear what the cause of the different rotational behavior among TO stars actually is (\citealt{2015Dantona}; \citealt{2020Bastian}; \citealt{2021Kamann}), while the predictions of the age scenario are at odds with observations. To explain the observed extension of MSTOs, younger and massive clusters should indeed experience star formation for the first few hundred million years of their lives, and they should be able to retain or accrete gas from their surroundings for a longer period of time (\citealt{2014Cabrera-ziri}, \citealt{2016Bastian}). For these reasons, it has also been suggested that none of these scenarios alone can explain the feature comprehensively (e.g., \citealt{correnti2017}, \citealt{Goudfrooij2017}
    and \citealt{2022Cordoni}). Furthermore, these scenarios may be characterized by additional complexity, such as the braking of some fast rotating stars, which would affect their evolution and their position at the eMSTO point \citep{2017Dantona}, or the role of circumstellar dust and mass loss from rotating stars in shaping the eMSTO \citep{2023Dantona}.
    
    \noindent An observable that can put important constraints on the theoretical models that try to explain the origin of the eMSTO is the radial distribution of its stars. In particular, in the age scenario, the younger population should generally have been born more centrally concentrated than the older one \citep{2011Goudfrooij}. Despite the numerous attempts at detecting differences in the radial distribution of red and blue eMSTO stars, conflicting results have been reported in the literature (e.g., \citealt{correnti2017}).
    \\
    In this context, an interesting object to study is the intermediate-age star cluster NGC~419 ($t \sim 1.5$ Gyr, \citealt{glatt2009}, $M \sim 2 \cdot 10^5 M_{\odot}$, \citealt{2014Goudfrooij}) situated in the Small Magellanic Cloud (SMC) at a distance $d \sim 59$ kpc \citep{2014Goudfrooij} and presenting a clear eMSTO (\citealt{2008Glatt}). As it was indeed proven in our previous study \citep{2022Dresbach}, this star cluster is dynamically young, which implies that the imprints of the formation of the populations describing the eMSTO have not yet been washed out by dynamical evolution \citep{dalessandro2019}. Additionally, the availability of multi-epoch observations from Hubble Space Telescope (HST) allowed us to obtain relative proper motion (PM) measurements for each star in the catalog, and thus to perform our analysis on a kinematically decontaminated cluster members population, which is particularly relevant for highly contaminated star clusters in the MCs (\citealt{massari2021}; \citealt{2022UnivMilone}, \citealt{2023milone}). When it was first discovered (\citealt{2008Glatt}), the eMSTO of NGC~419 was attributed to a prolonged period of star formation ($\sim 700$ Myr, \citealt{Rubele2010}). In addition, the CMD of this cluster presented a secondary red clump (RC, \citealt{1999Girardi}), a rare photometric feature that was attributed to a population of younger, more massive stars by \cite{Girardi2009}, who were hence able to reproduce both features by assuming an age difference. However, in a recent spectroscopic study of the cluster TO stars, \cite{2018Kamann419} detected a difference in the rotational velocity values among blue (slow) and red (fast) stars, which would contradict the previous hypothesis. Additionally, \cite{2016WuLi} suggest that the morphology of the NGC~419 subgiant branch (SGB) is not compatible with an internal age spread, but it could rather be explained with the rotation scenario. 
    Different rotation velocities could thus partially explain NGC\,419 features, but other mechanisms might also be at play (\citealt{2018Marino}; \citealt{2018Kamann419}), which a conjoint analysis of eMSTO and eRC might better reveal.
    \\
    To provide some clarity on these results, in this work we analyze the radial distributions of stars populating the eMSTO and the extended red clump (eRC), in search for different levels of segregation among their subpopulations. Radial distributions have proven to be a valuable tool to understand the origin of these features (e.g., \citealt{2011Goudfrooij}, \citeyear{2014Goudfrooij}), and computing them for a sample of kinematically selected member stars of NGC~419 offers a unique opportunity to robustly test the physical processes behind their formation.
    \\
    The outline of the paper is the following: in Section \ref{sec:data}, we present the data analysis and the decontamination procedure. In Section \ref{sec:radial}, we build and examine the radial distributions of the different populations of the eMSTO and eRC stars. In Section \ref{sec:discus} we investigate the different scenarios that could explain the features of NGC~419 and discuss our results. A summary is provided in Section \ref{sec:conclusio}.
\section{Data analysis}\label{sec:data}
\begin{figure}[h]
\centering
\includegraphics[scale=0.365]{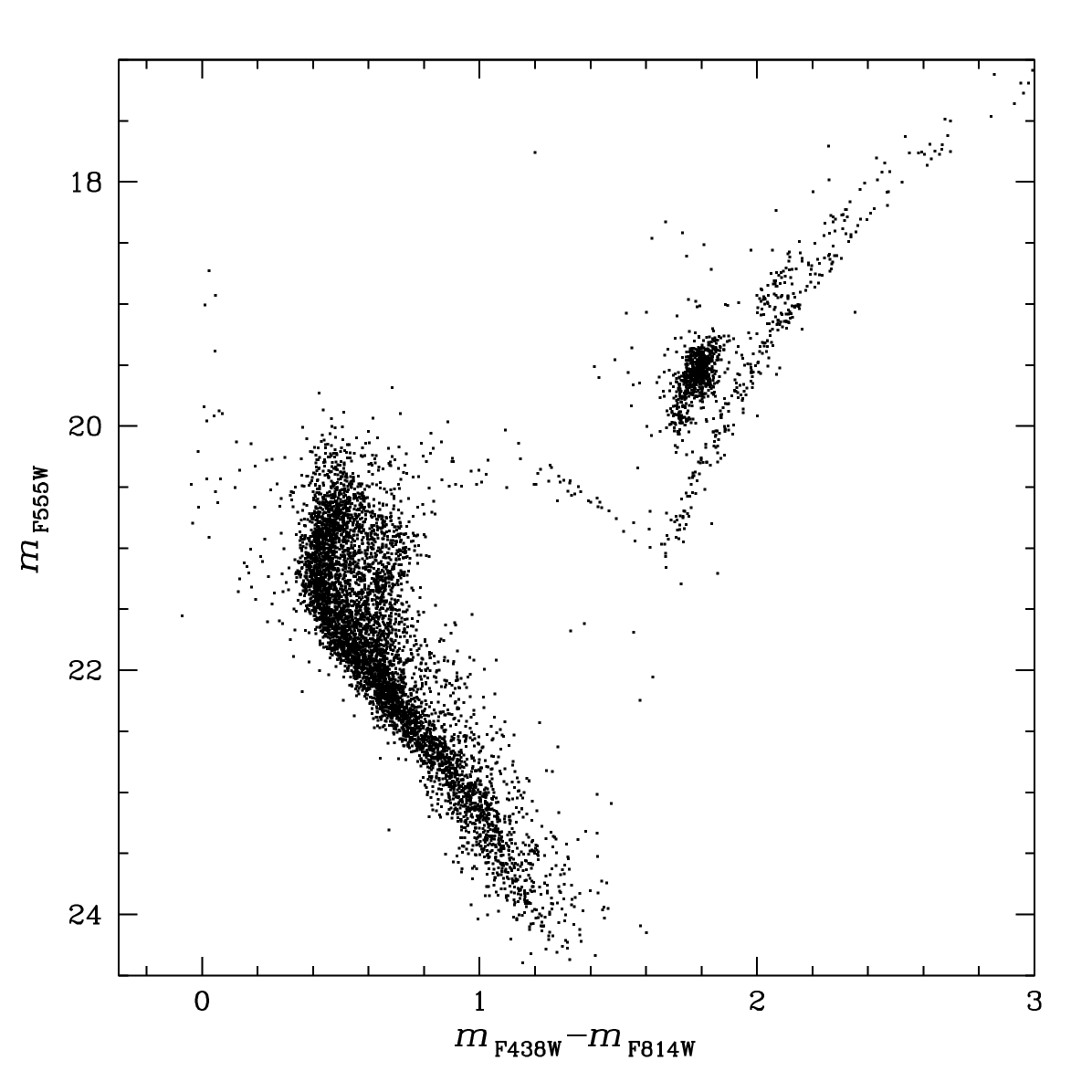}
\caption{CMD of NGC\,419 resulting from both the PM-based membership selection and the quality selection.}
\label{fig:2sigma419}
\end{figure}

    The photometric data for NGC~419 were acquired through observations conducted using the HST. The Ultraviolet–Visible Channel (UVIS) of the Wide Field Camera 3 (WFC3) was used to acquire the images in the F336W and F438W filters, while the images in the F555W and F814W filters were obtained with the Advanced Camera for Surveys (ACS) Wide Field Channel (WFC).
    The detailed list of observations can be found in Table 1 of \cite{massari2021}\footnote{The data used can also be found in MAST and accessed via \href{https://dx.doi.org/10.17909/g6s0-yr36}{DOI:10.17909/g6s0-yr36}}. 
    Thanks to the long temporal baselines provided by these multi-epoch observations, of about 10.75 yr, measurement of the relative PMs were performed by \citeauthor{massari2021} (\citeyear{massari2021}; see also the description in \citealt{2022Dresbach}). This is why only a brief description of the adopted procedures is provided here (readers can refer to those papers for details about the photometric analysis and PM derivation). To remove cluster contamination from field stars (Milky Way + SMC) and select cluster members, we took advantage of the kinematical measurements, plotted on a vector point diagram (VPD). We indeed considered as cluster members only the stars within a radius  that is double the expected total dispersion of the cluster population ($\sigma$) from the origin of the VPD.
    \begin{figure}[h]
\centering
\includegraphics[scale=0.365]{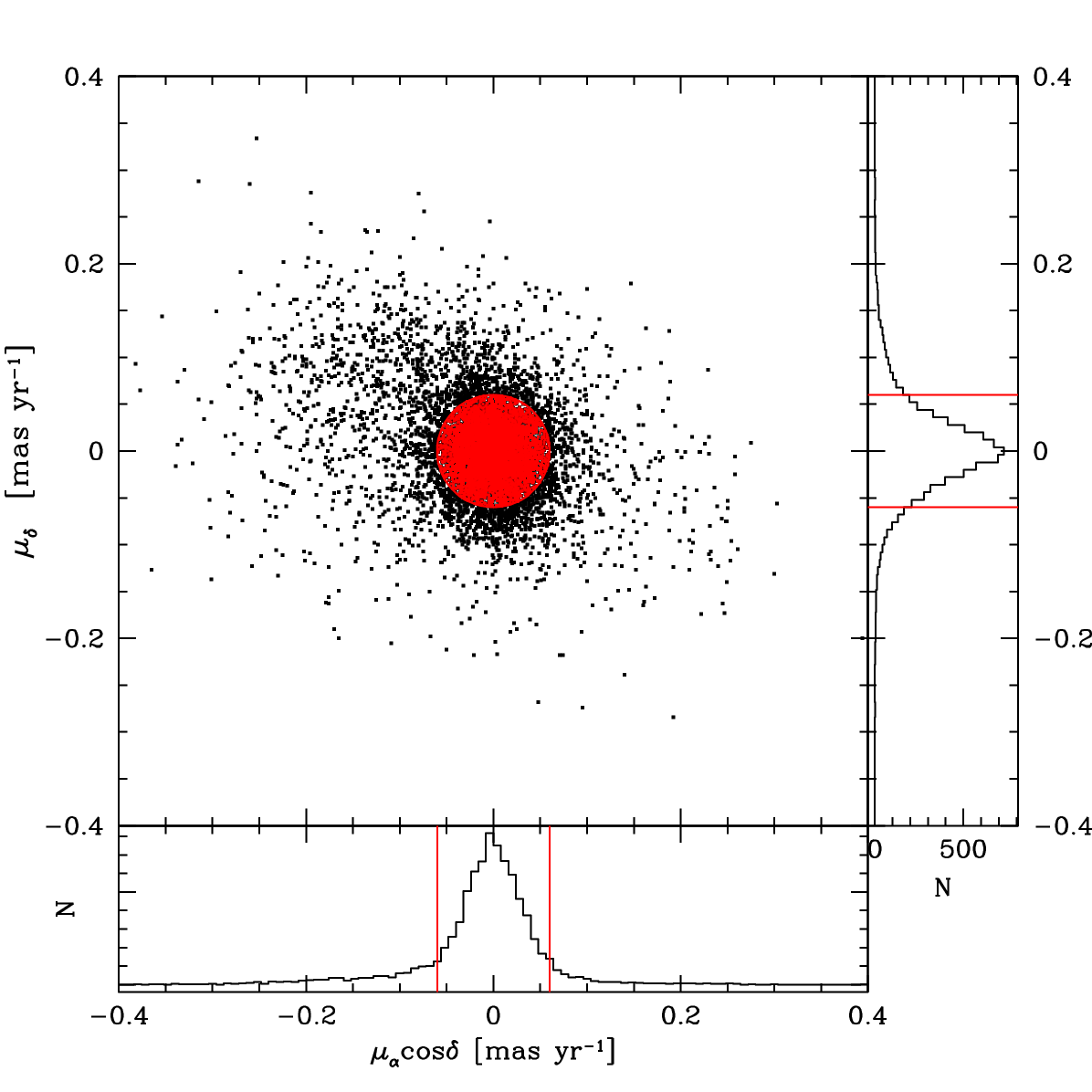}
\caption{ VPD of bright sources in the catalog ($m_{\rm F555W}<22.3$, in black) and the stars selected as cluster members (in red). The histograms in the bottom and right panels show the distribution of the bright sources in the R.A. and Dec. PM component, respectively.}
\label{fig:VPD}
\end{figure}
    The location in the VPD of the likely members selected is shown in Fig. \ref{fig:VPD} (red dots).
    The value of the dispersion $\sigma$ was determined by two independent terms, the intrinsic velocity dispersion of the cluster and the error associated with the PM measurement, summed in quadrature (\citealt{2022Dresbach}). The resulting value for the total dispersion is $\sigma = 0.03 \  \rm{mas \ yr^{-1}}$.  As we discuss in \cite{2022Dresbach}, we adopted a $2\sigma$ selection because it represented the best compromise between the number of members selected and the residual contamination included in the selection. A quality selection was performed on the photometric and astrometric catalogs, aimed at removing poorly measured stars. The adopted criteria are listed in \cite{2022Dresbach} and are based on the photometric error, PM errors, $\chi^2$ values of the PM analysis, the \texttt{RADXS} parameter, and the number of detections for each source. The CMD of the cluster resulting from both the PM-based membership selection and the quality selection is shown in Fig. \ref{fig:2sigma419}.

\section{Radial distributions}\label{sec:radial}
\begin{figure}[h]
\centering
\includegraphics[scale=0.365]{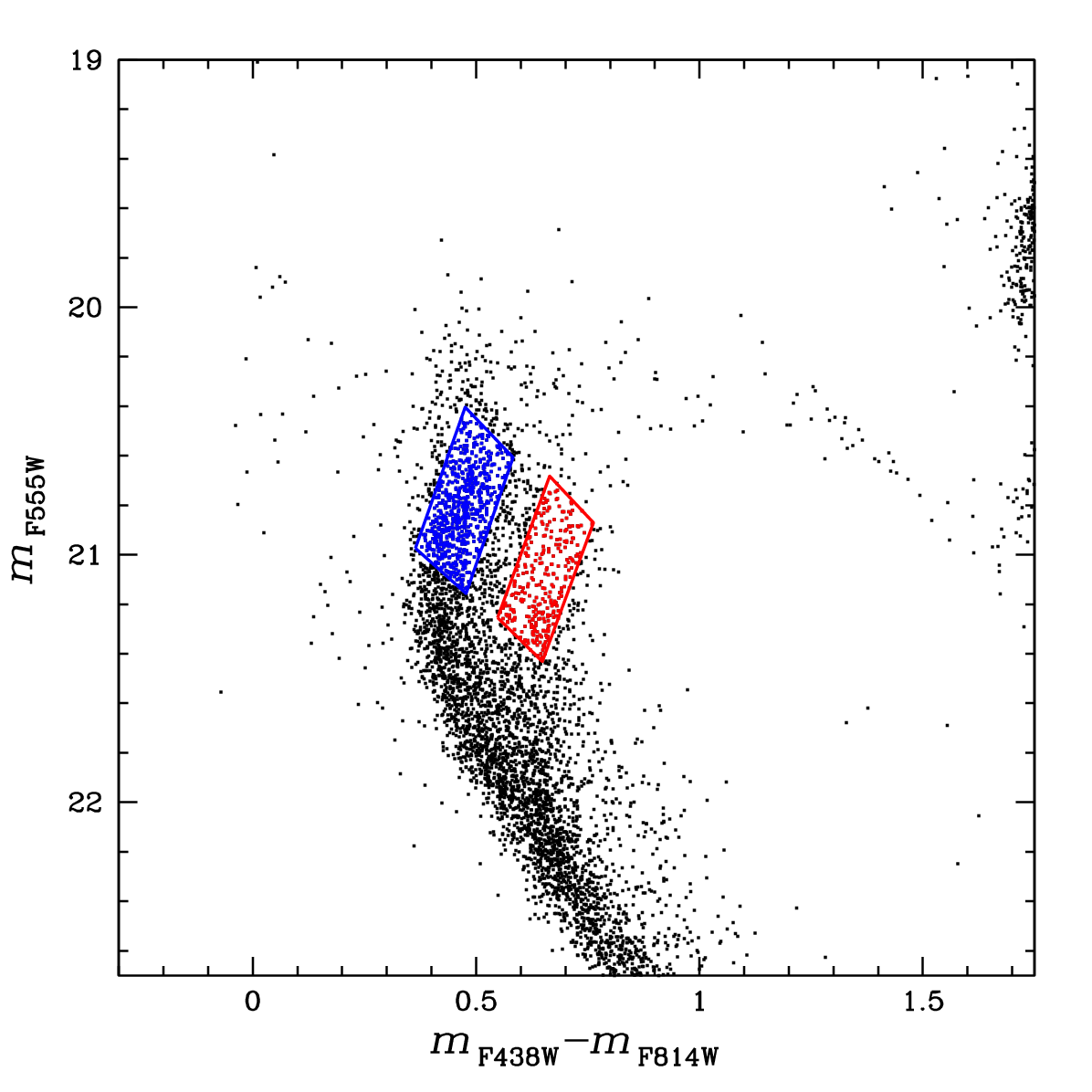}
\caption{CMD of the eMSTO of NGC\,419. The subpopulations selected for the analysis are shown in blue and red.}
\label{fig:selTO}
\end{figure}

\begin{figure}[h]
\centering
\includegraphics[scale=0.365]{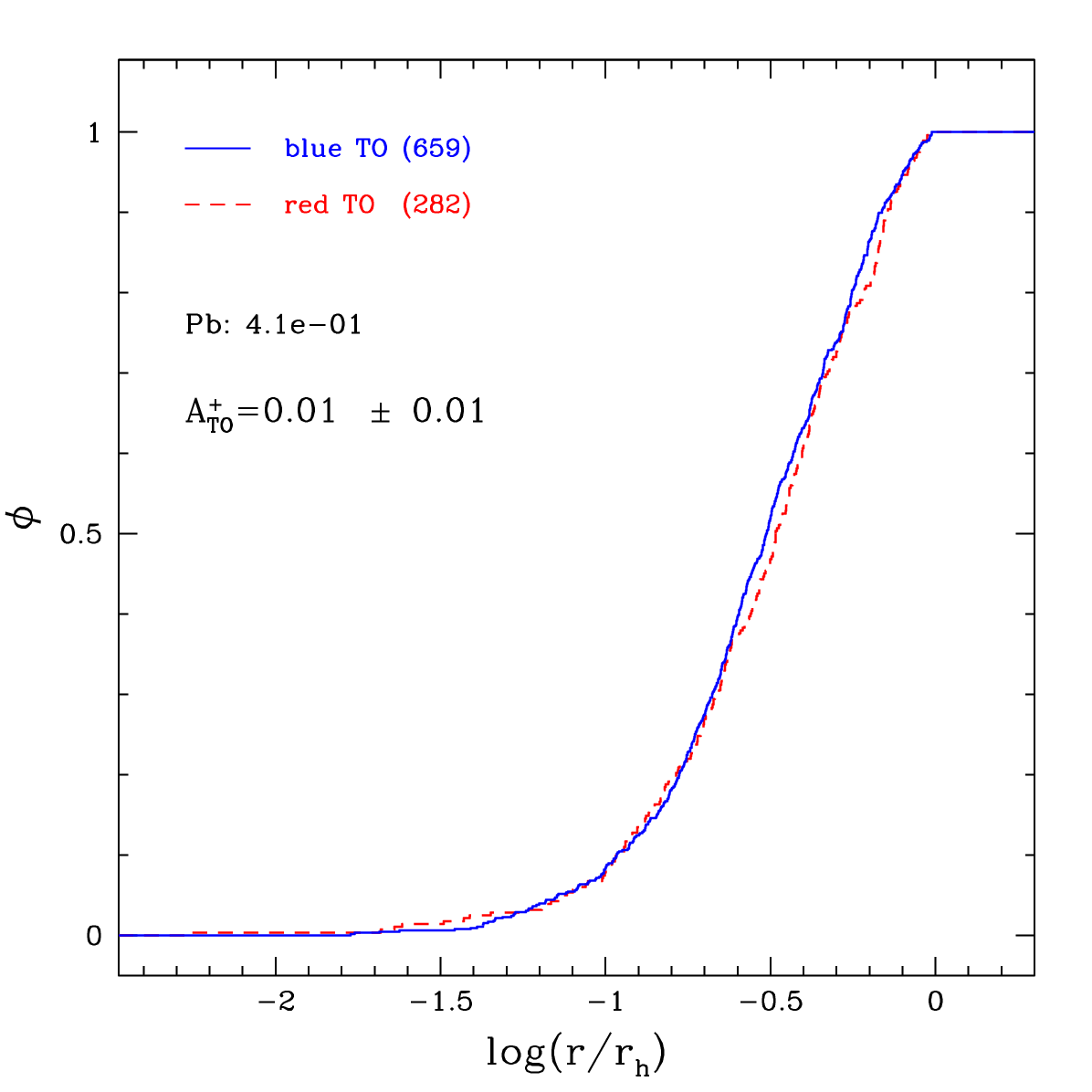}
\caption{Cumulative radial distributions of the two TO populations selected in Fig. \ref{fig:selTO}. We also marked the Kolmogorov-Smirnov probability that the two distributions were extracted from the same parent family (Pb) and the value of the $A^{+}$ parameter ($A^{+}_{\rm TO}$).}
\label{fig:apiuTO}
\end{figure}

As a way to unravel the nature of the eMSTO in NGC~419, we analyzed the radial distributions of eMSTO and eRC subpopulations to determine whether they present some difference in their level of segregation. To avoid possible issues caused by different levels of completeness, we only compared features with comparable levels of magnitude. To quantify the segregation, we used the $A^{+}$ parameter (originally introduced for the study of blue straggler stars, see \citealt{alessandrini2016}; \citealt{Lanzoni2016}; \citealt{ferraro2018}, \citeyear{FERRARO2019}, \citeyear{2023Ferraro}; \citealt{2022Dresbach}), which measures the area enclosed between two cumulative radial distributions of stars, $\phi_{\alpha}(x)$ and $\phi_{\beta}(x)$:

\begin{equation}
    A^{+}(x)=\int_{x_{\rm min}}^x \left (\, \phi_{\alpha}(x') -\phi_{\beta}(x') \, \right ) \, dx'
,\end{equation}

\noindent where $x={\rm log} ( r/r_{\rm h})$ is the cluster-centric distance normalized to the half-mass radius ($r_{\rm h}=36\farcs 73$; \citealt{2022Dresbach}) and expressed in logarithmic units. The minimum value sampled is denoted by $x_{\rm min}$ (often chosen to coincide with the cluster center\footnote{The position of the photometric center of the cluster adopted for the analysis is the one derived by \cite{glatt2009}: $\alpha = 1^h 08^m 17\fs31$  and $\delta = - 72^{\circ} \, 52 \arcmin \, 02\farcs 49$.}, where its value corresponds to zero). For this study, we extended the analysis to twice the half-mass radius (hence, $x=2$). With this definition, the value of $A^{+}$ increases with the level of segregation of population $\alpha$ with respect to that of the population $\beta$.
To determine the value of $A^{+}$, we first needed to select the populations we wanted to analyze from the CMD. We started our analysis by sampling the eMSTO in its extreme blue and red edges, as shown in Fig. \ref{fig:selTO}. 
For the computation of $A^{+}$, we considered the stars in the blue edge as population $\alpha$, and the stars in the red edge as population $\beta$.
We then built the cumulative radial distributions of these stars within two half-mass radii and computed the value of $A^{+}_{\rm TO}$ and the associated uncertainty, obtained by applying a jackknife bootstrapping technique (\citealt{Lupton93}). Additionally, we performed a Kolmogorov–Smirnov (KS) test aimed at determining whether the two populations can be considered statistically different, and at which level. The results are shown in Fig. \ref{fig:apiuTO}. From this plot, it is clear that the distributions are almost identical, with the value of the $A^{+}$ parameter being fully consistent with zero: $A^{+}_{\rm TO}= 0.01 \pm 0.01$. This means that eMSTO stars are equally segregated regardless of their color. The results of the KS test indicate that the two spatial distributions are statistically identical, which is consistent with the measurement of $A^{+}_{\rm TO}$. This finding supports the stellar rotation scenario, where no difference in segregation is expected among TO stars and argues against the age scenario.
To check the robustness of this conclusion, we investigated whether our final results were affected by the selection processes, using the methodology also applied in \cite{2022Dresbach}. We thus estimated new values of $A^{+}$ on a catalog of NGC~419 stars that were selected solely with the $2\sigma$ kinematic selection method and not based on the quality criteria detailed in Section \ref{sec:data}. 

\noindent A new estimate for the value of $A^{+}_{\rm TO}$ was obtained, $A^{+}_{\rm TO} = 0.02 \pm 0.02 $, consistent with the original estimate ($A^{+}_{\rm TO}=0.01 \pm 0.01$). This provides evidence that the final result was not substantially impacted by the procedure implemented to eliminate low-quality measurements. Similarly, when we only applied the quality criteria and did not consider the kinematic selection, we obtained the new value of $A^{+}_{\rm TO}= 0.01 \pm 0.02$, which is still consistent with our results. 

\noindent NGC\,419 also presents a secondary RC that is fainter than the main one and located in the CMD of Fig. \ref{fig:2sigma419} at $m_{\rm{F555W}}=20$ and $(m_{\rm{F438W}}-m_{\rm{F814W}})=1.7$, which has been interpreted by \cite{Girardi2009} as due to an age difference, thus allowing us to test the age scenario directly.
\begin{figure}[h]
\centering
\includegraphics[scale=0.365]{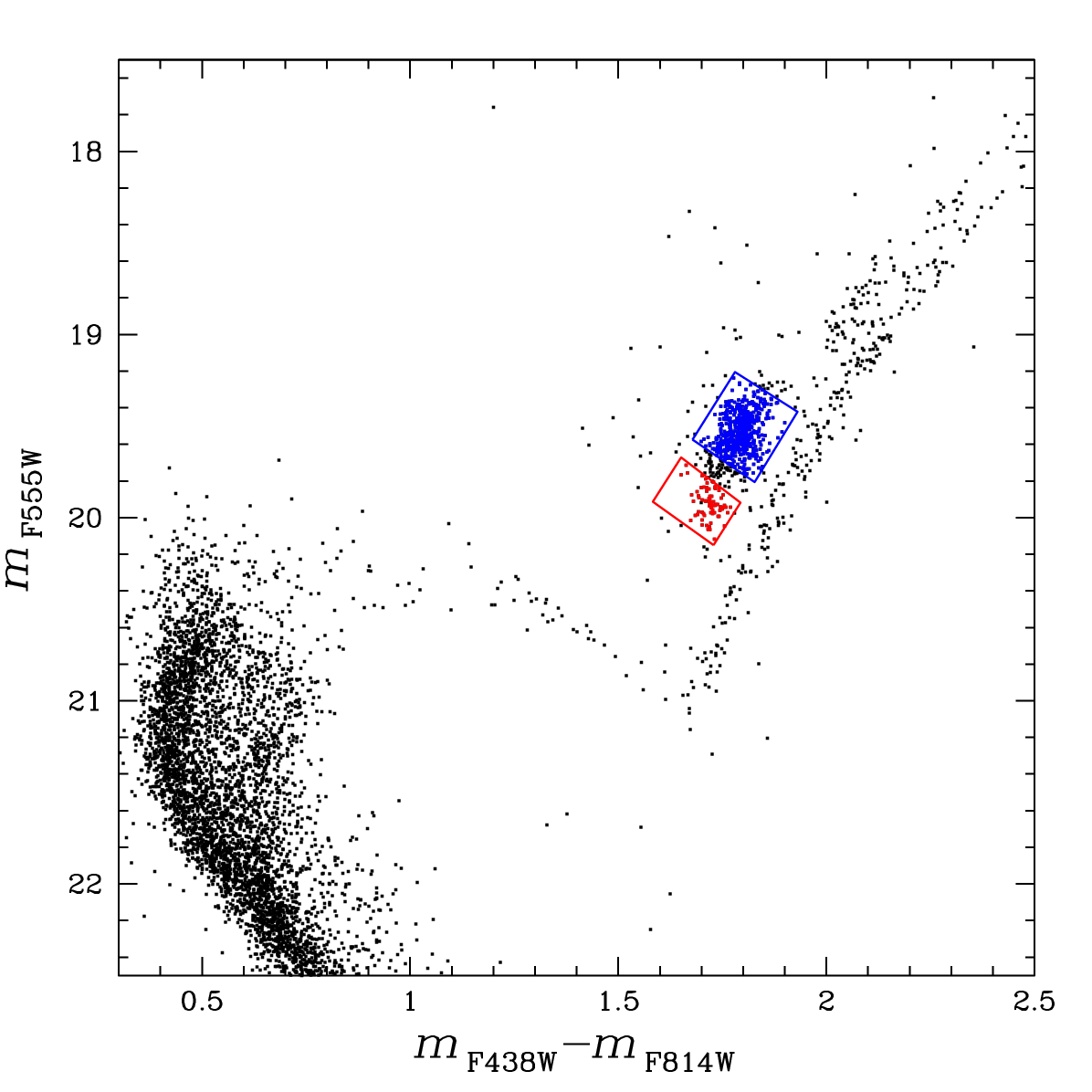}
\caption{CMD of the eRC of NGC\,419.  The subpopulations selected for the analysis of the bright and faint extension of the RC are shown in blue and red, respectively.}
\label{fig:RCselect}
\end{figure}
We thus selected the two populations, as shown in Fig. \ref{fig:RCselect}, by separating the primary clump from its extension toward fainter magnitudes. We determined the parameter $A^{+}_{\rm RC}$ based on the cumulative radial distributions for the two populations within two $r_{\rm h}$ (Fig. \ref{fig:apiuRC}) and obtained a value of $A^{+}_{\rm RC}=-0.04 \pm 0.04$, which is consistent with zero at a 1 sigma uncertainty. This means that, consistent with our results for the eMSTO analysis, there is no statistically significant difference in the level of segregation of the two RC populations. We again studied the effects of the selection procedures on these final results. We applied the same analysis performed for the eMSTO and found values consistent with our initial results ($A^{+}_{\rm RC}= - 0.03 \pm 0.04$ when we removed the kinematic decontamination and $A^{+}_{\rm RC} = -0.01 \pm 0.04 $ when we did not apply the quality selection). It should indeed be noted that the contamination by SMC or Milky Way field stars at the RC level is negligible, and this is why $A^{+}$ is so scarcely affected when ignoring the membership selection.

\noindent To test the robustness of these results, we repeated the analysis by using different methods to select the subpopulations at the MSTO and RC level, as well as to estimate their level of segregation. For the latter, we employed the method described in \cite{2014Libralato}, which estimates the level of segregation between the subpopulations by first fitting the color (and magnitude) distribution of stars along the eMSTO (and eRC) with bi-Gaussian functions at different radial bins and then comparing the areas of each of the two Gaussians measured at different radial bins to detect any variation. We always found consistency with the results presented in this section.

\begin{figure}[h!]
\centering
\includegraphics[scale=0.365]{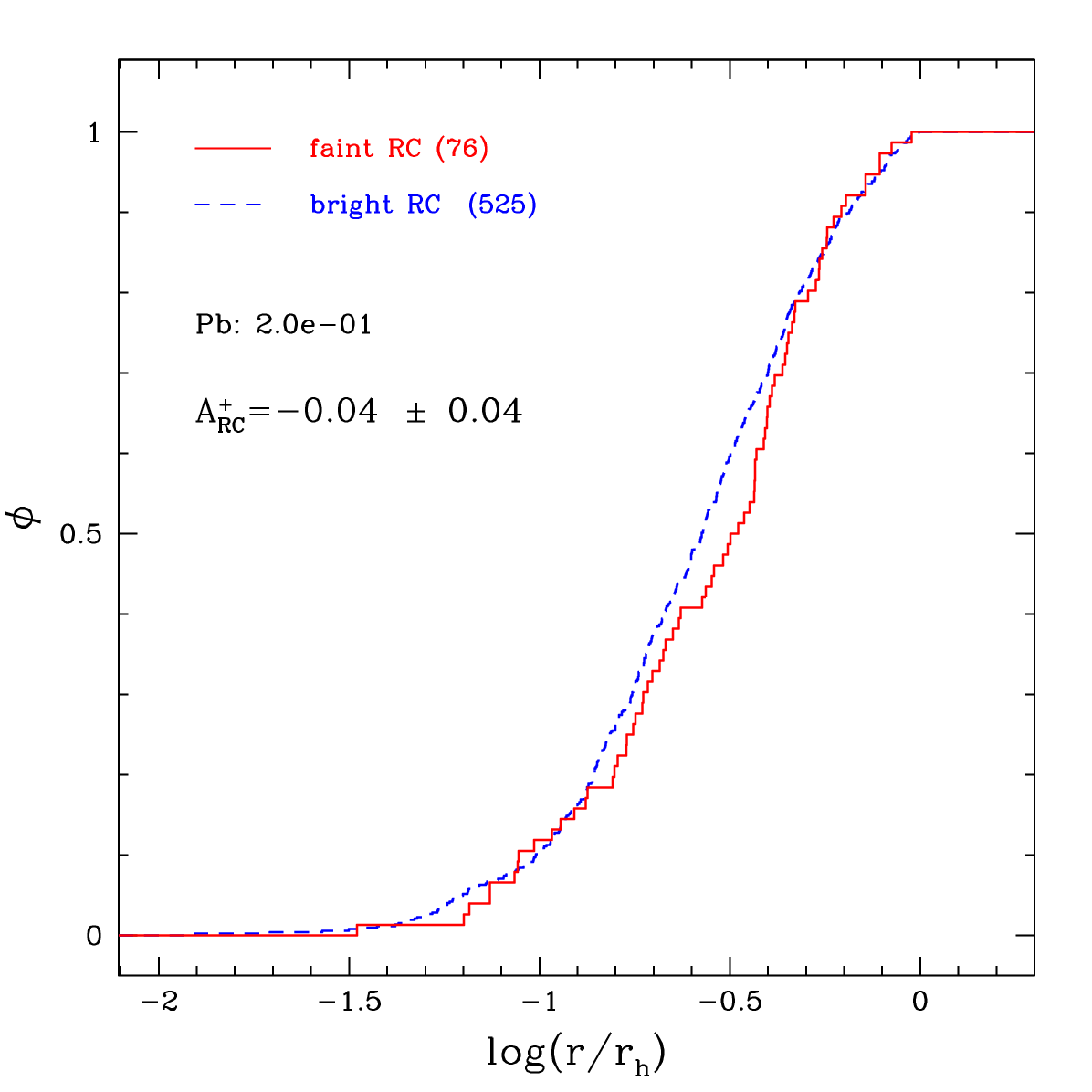}
\caption{Cumulative radial distributions of the two RC populations selected in Fig. \ref{fig:RCselect}. We also marked the Kolmogorov-Smirnov probability that the two distributions were extracted from the same parent family (Pb) and the value of the $A^{+}$ parameter ($A^{+}_{\rm RC}$).}
\label{fig:apiuRC}
\end{figure}

\section{Discussion and interpretation}\label{sec:discus}
As we introduced in Section \ref{sec:intro}, a few different scenarios have been proposed to explain the peculiar TO and RC features. Here we discuss them in light of our findings.

\begin{figure}[h!]
\centering
\includegraphics[scale=0.365]{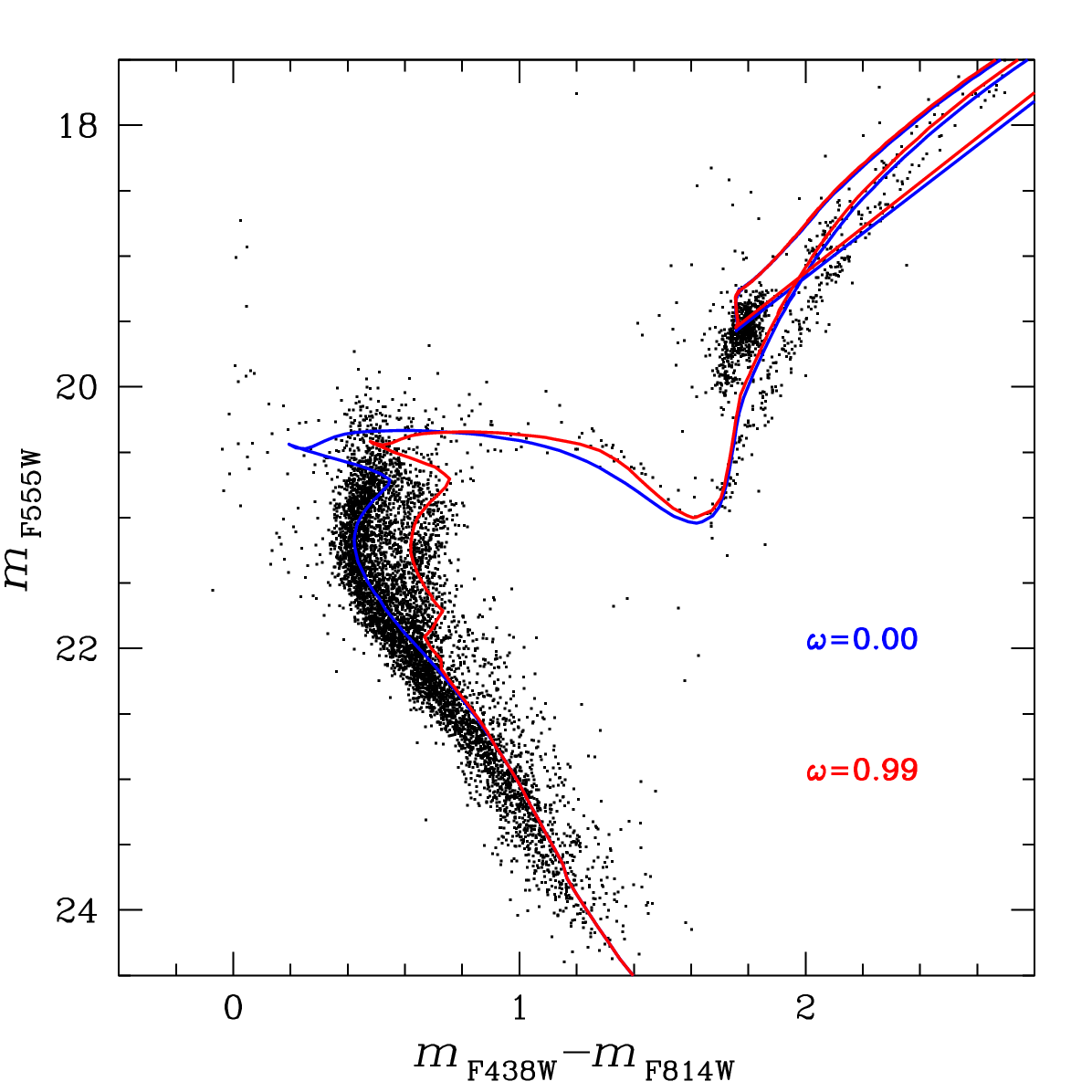}
\caption{Isochrones built with \textsc{parsec} evolutionary models for stars with different rotational rates (see labels) and age $t_{\rm age} = 1.5$ Gyr, superposed to the observed CMD.}
\label{fig:isoRotaz}
\end{figure}

\noindent Our analysis on the radial distribution of TO and RC stars shows that there is no statistically significant difference in the level of radial segregation of stars populating different regions of these extended CMD features. Therefore, both of these results would exclude the age spread scenario, which attributes the blue extension of the MSTO and the fainter portion of the RC to younger stars. The reason is that according to this scenario, the younger population should form more segregated toward the cluster center and, since the system is still dynamically young \citep{2022Dresbach}, we should detect a difference between the two radial distributions. Moreover, to this day, no other evidence of multiple populations has been found for this cluster (\citealt{2017Martocchia}, \citealt{2020Cabrera}). Therefore, our findings support the idea that at least the eMSTO feature is caused by stars with different rotational velocities. This interpretation is primarily supported by the results of \cite{2018Kamann419}, who detected a difference in the rotational velocities of TO stars, as well as by the analysis of \cite{2016WuLi} on the “converging” morphology of the cluster SGB, whose tightness cannot be explained with the age scenario. \\
Nevertheless, by invoking stellar rotation, it is still difficult to explain the origin of the secondary RC since high rotational velocities are not expected for evolved stars, which should lose angular momentum due to their increase in radius. Additionally, isochrones built using stellar rotation models fail to reproduce the eRC. In Fig. \ref{fig:isoRotaz} we plotted isochrones built using the new \textsc{parsec} evolutionary models \citep{2022Nguyen} and computed for an age of $t_{\rm age} = 1.5$ Gyr and a metallicity of $Z=0.004$, considering two different rotational rates\footnote{The rotation rate $\omega$ is the equatorial angular velocity of the star $\Omega$, divided by its critical value $\Omega_{\rm crit}$} $\omega=0.0$ and 0.99. While the two isochrones were able to reproduce the color spread at the MSTO level, no differences are predicted for the RC. 

\begin{figure}[h!]
\centering
\includegraphics[scale=0.365]{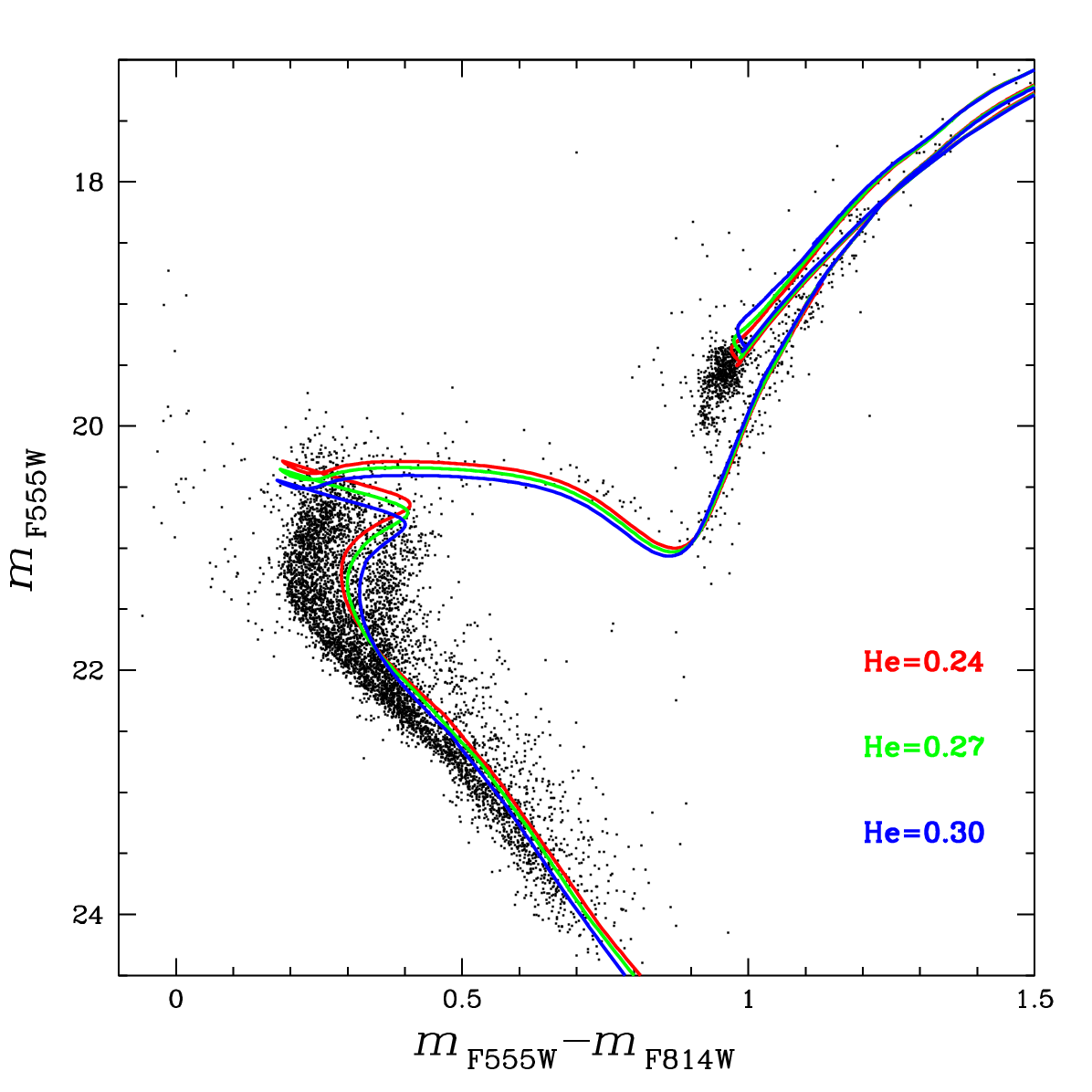}
\caption{Isochrones built with BaSTI evolutionary models for nonrotating stars with the same age and metallicity but different He abundances (see labels), superposed to the observed CMD.}
\label{fig:CMDelio}
\end{figure}

\noindent Given this apparent contradiction between our findings and model predictions, we explored alternative explanations for what we observed in NGC~419. One possibility is an enhanced He abundance. To investigate it, we built isochrones for nonrotating stars using the BaSTI database\footnote{\url{http://albione.oa-teramo.inaf.it}} \citep{2006Pietrinferni}. We assumed the same age as in the previous model, metallicity $\rm{[Fe/H]=-0.7}$, and three different He abundances (${\rm Y} = 0.24$, 0.27, and 0.30), and we adopted alpha-enhanced models for the heavy elements mixture. These three models are shown in Fig. \ref{fig:CMDelio}, where we can see that a difference in the He abundance cannot reproduce the observed extension of TO and RC.

\noindent A further, more exotic mechanism that, in principle, could explain the shape of the RC is a differential mass loss during the red giant branch (RGB) phase, in combination with the different rotational velocities at the TO. If stars with the same TO mass and the same age undergo a different mass loss during their RGB evolution, they reach the RC with a different mass. To test this hypothesis, we built BaSTI evolutionary tracks with different initial masses and found that the RC morphology could only be reproduced with a difference in the internal mass loss of $\Delta M > 0.4 \rm{M_{\odot}}$. Such a value is extremely unlikely (e.g., \citealt{2021MarcoTailo}) and leads us to exclude this scenario. However, in their analysis on NGC~419, \cite{Girardi2009} were able to reproduce the RC morphology using PARSEC models \citep{2012Bressan} with a mass difference of around $0.1 \rm{M_{\odot}}$ (corresponding to an age difference of 300 Myr).
This could indicate that the difficulties in reproducing both the eMSTO and the eRC without invoking any age difference might be due to shortcomings in the stellar models. Additionally, when comparing the numerical ratios of bright and faint RC, red and blue TO, it appears, counterintuitively, that the stars that would experience the stronger mass loss (meaning those located along the more populated bright portion of the RC) should be those that rotate slowly or do not rotate at all (meaning those located along the more  populated blue portion of the TO). This would suggest that the reason for the enhanced mass loss is not related to stellar rotation, which would imply a "nonstandard" mass-loss mechanism. Indeed, by looking at other intermediate-age clusters coeval to NGC~419, we noticed that while the majority of them are characterized by an eMSTO, only a few also present an eRC (\citealt{girardi2013}; \citealt{2011rubele}, \citeyear{2013Rubele}, \citealt{2014Goudfrooij}). This evidence might support, or at least does not discredit, this nonstandard mechanism of mass loss, but in this case we lack the instruments to characterize and discuss it in more detail.

In conclusion, the difficulty encountered in reproducing the morphology of the eMSTO and the eRC simultaneously leaves us with two possible interpretations. Either different mechanisms are shaping the two CMD features, or the available theoretical models still have shortcomings that prevent them from successfully reproducing both features.


\section{Summary}\label{sec:conclusio}
In this paper we analyzed multi-epoch HST observations in the direction of the intermediate-age star cluster NGC~419 situated in the SMC.
With the aim of testing the different interpretations regarding its CMD features, we analyzed the radial distributions of eMSTO and eRC subpopulations to determine whether they present some differences in their level of segregation, as a tool to understand their origin. This work is the first of its kind performed on the eMSTO and eRC of a dynamically young cluster \citep{2022Dresbach}, where the kinematical properties of the different stellar populations have not yet been washed out by dynamical evolution. Thanks to the availability of an astrometric dataset, the catalog was first kinematically decontaminated from field stars. The different subpopulations were then selected based on the positions of stars in the cluster CMD, and their cumulative radial distributions were built.
By computing the $A^{+}$ parameter to measure the relative radial segregation of the populations observed along the red and blue edges of the eMSTO, we detected no difference. No segregation was detected when comparing the bright and faint stars along the eRC either, despite being previously attributed to a prolonged period of star formation \citep{Girardi2009}. These results would exclude the presence of an age spread among TO and RC stars, while instead supporting the rotation scenario as a possible interpretation, where no difference in the radial distributions is expected. 
Our analysis is in agreement with results from previous works (\citealt{2018Kamann419}, \citealt{2016WuLi}) ascribing the origin of the eMSTO to a spread in rotational velocities and excluding the age spread scenario.
Nonetheless, stellar evolution models still fail to explain the presence of an eRC with a rotational velocity spread, and opens the possibility that the eRC has originated from a different process. In this regard, we excluded helium variations as an alternative explanation for the secondary RC. Lastly, as a further, more exotic scenario able to explain our observational findings, we propose the possibility that a nonstandard mechanism of differential mass loss along the RGB phase could have taken place and shaped the morphology of the RC, whose origin might then be unrelated to the eMSTO feature. Shortcomings in the evolutionary models might also explain the apparent failure at reconciling all the observed properties of TO and RC stars.

\begin{acknowledgements}  
    We thank the anonymous referee for comments and
suggestions that improved the quality of our paper.
    This research is part of the project Cosmic-Lab (“Globular Clusters as Cosmic Laboratories”) at the Physics and Astronomy Department of the Bologna University (see the web page: \url{http://www.cosmic-lab.eu/Cosmic-Lab/Home.html}). The research is funded by the project Light-on-Dark granted by MIUR through PRIN2017K7REXT contract (PI: Ferraro). This work is also based on observations made with the NASA/ESA Hubble Space Telescope, obtained from the Data Archive at the Space Telescope Science Institute, which is operated by the Association of Universities for Research in Astronomy, Inc., under NASA contract NAS 5-26555.
\end{acknowledgements}



\bibliographystyle{aa}
\bibliography{Bibliography}

\end{document}